\documentclass[%
reprint,
amsmath,amssymb,
 ]{revtex4-1}

\usepackage{graphicx}
\setcitestyle{super}

\begin{document}

\title{\Large\bfseries\noindent\sloppy \textsf{Photo-acoustic dual-frequency comb spectroscopy}}

\author{Thibault~Wildi$^{1,2}$}
\thanks{These authors contributed equally.}
\author{Thibault~Voumard$^{1,2}$}
\thanks{These authors contributed equally.}
\author{Victor~Brasch$^{1}$}
\author{Gürkan~Yilmaz$^{1}$}
\author{Tobias~Herr$^{1,2}$}%
\email{tobias.herr@cfel.de}
\affiliation{
$^1$Swiss Center for Electronics and Microtechnology, Rue de l'Observatoire 58, 2000~Neuchâtel, Switzerland\\
$^2$Center for Free-Electron Laser Science, Deutsches Elektronen-Synchrotron, Notkestr. 85, 22607 Hamburg, Germany
}

\maketitle

\textbf{Photo-acoustic spectroscopy (PAS) is one of the most sensitive non-destructive analysis techniques for gases, fluids and solids. It can operate background-free at \textit{any} wavelength and is applicable to microscopic and even non-transparent samples. Extension of PAS to broadband wavelength coverage is a powerful tool, though challenging to implement without sacrifice of wavelength resolution and acquisition speed.
Here, we show that the unmatched precision, speed and wavelength coverage of dual-frequency comb spectroscopy (DCS) can be combined with the advantages of photo-acoustic detection. Acoustic wave interferograms are generated in the sample by dual-comb absorption and detected by a microphone. As an example, weak gas absorption features are precisely and rapidly sampled; long-term coherent averaging further increases the sensitivity. This novel approach of \textit{photo-acoustic dual-frequency comb spectroscopy} generates unprecedented opportunities for rapid and sensitive multi-species molecular analysis across all wavelengths of light.
}  

\begin{figure}
    \includegraphics[width=\columnwidth]{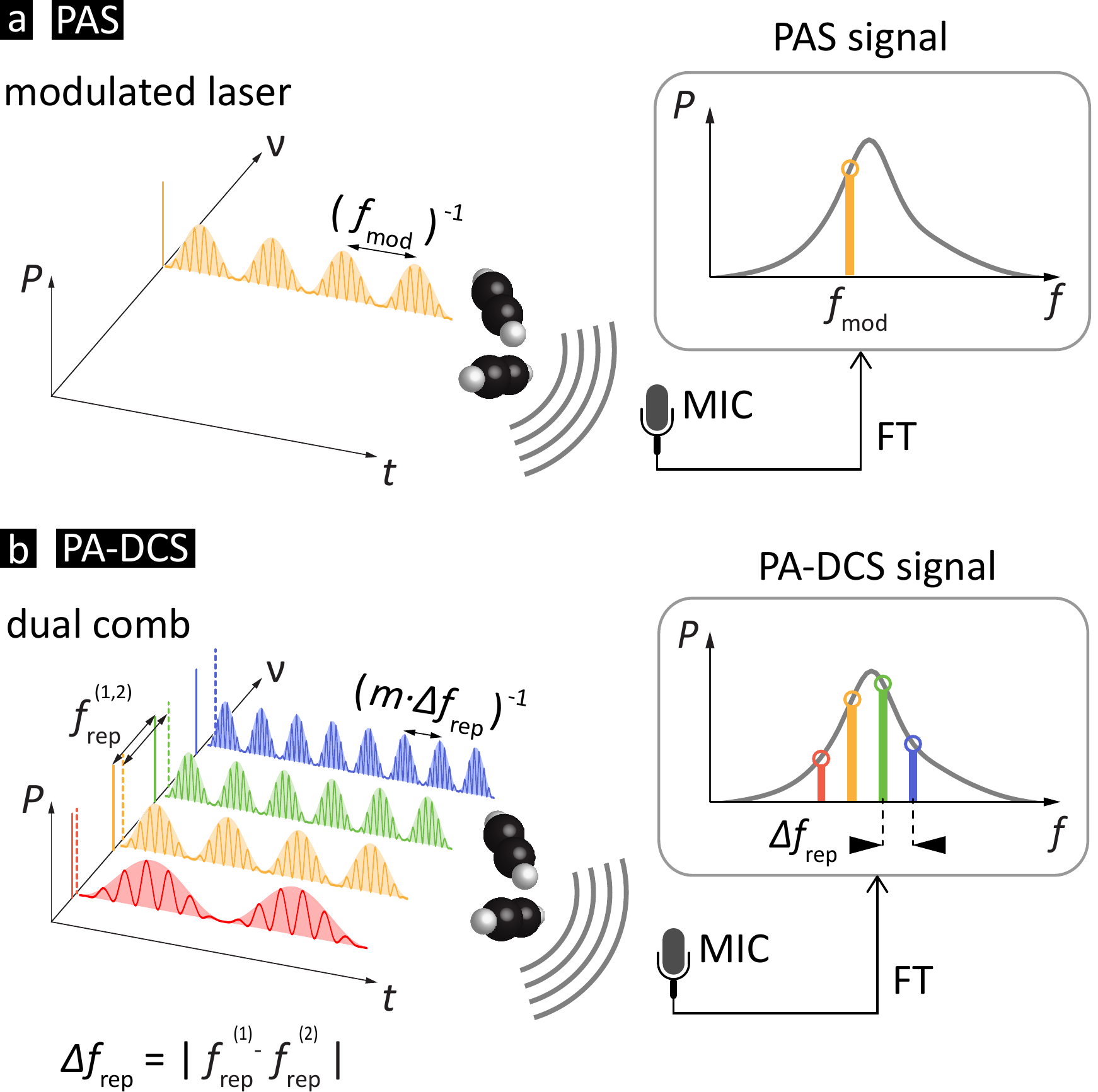}
    \caption{\textbf{Photo-acoustic dual-comb spectroscopy.} 
        \textsf{\textbf{a}},~In photo-acoustic spectroscopy (PAS), absorption of a modulated laser results in acoustic waves that are recorded by a microphone (MIC). The acoustic spectrum (after Fourier transformation, FT) contains the PAS signal tone at the modulation frequency $f_\mathrm{mod}$ that indicates the strength of the optical absorption. \textsf{\textbf{b}},~Photo-acoustic dual comb spectroscopy (PA-DCS) uses broadband dual frequency combs, whose repetition rates $f_\mathrm{rep}^{(1)}$ and $f_\mathrm{rep}^{(2)}$ differ by a small amount $\Delta f_\mathrm{rep}$. The PA-DCS signal is comprised of multiple heterodyne acoustic tones that simultaneously sample the optical absorption spectrum at multiple frequencies. ($P$: power; $\nu$ and $f$: optical and acoustic frequencies; $t$: time.)
        }
        \label{fig1}
\end{figure}

\smallskip
In PAS\cite{kreuzer1971,rosencwaig1973, patel1979, li2011} , optical absorption of a modulated light source leads to periodic heating of a sample and the generation of an acoustic wave that can be detected by a microphone or an equivalent transducer (Figure~\ref{fig1}a). As the detection relies on the acoustic waves (rather than a weak attenuation of an optical signal), photo-acoustic detection can be background-free, with high signal-to-noise ratio (SNR), and importantly, works at \textit{any} wavelength of light. These unique properties have established PAS in environmental studies, solid state physics, chemical process control, medical application and life science, including for instance absorption measurements in atto-liter droplets \cite{cremer2016}, real-time monitoring of an ant's respiration \cite{herpen2006} and in-vivo tomographic imaging \cite{wang2003}. Quartz-enhanced photo-acoustic spectroscopy (QEPAS)\cite{kosterev2002, wu2017} and cantilever-enhanced photo-acoustic spectroscopy (CEPAS) \cite{koskinen2007, peltola2013} have enabled ultra-sensitive trace gas detection below the part-per-trillion-level\cite{spagnolo2012,tomberg2018}. %
Usually, PAS is performed at one single probing laser wavelength not ideal for the study of multiple species or studies in the presence of uncontrolled background absorption. Multiple laser sources can alleviate this problem to some extent, however, remain constraint to specific use cases. Therefore, in order to achieve broadband wavelength coverage, photo-acoustic detection has been combined with a Fourier-transform infrared spectrometer (FTIR-PAS) \cite{busse1978}. The achievable resolution is determined by the scan range of the interferometer, which can reach several meters for high resolution instruments. In addition to incoherent light sources, such as super-continua \cite{mikkonen2018}, coherent broadband spectra (unresolved optical frequency combs) have also been used to improve the overall performance \cite{sadiek2018, karhu2019}, and might allow adopting techniques for sub-nominal resolution \cite{maslowski2016} in the future. As such FTIR-PAS represents a powerful tool for broadband photo-acoustic spectroscopy, however, its resolution and acquisition speed remain limited due to the mechanical scan of the interferometer.\\

In this work, we show that the unmatched precision,  speed  and  wavelength  coverage  of  dual-frequency  comb  spectroscopy  (DCS)\cite{coddington2016, picque2019, schiller2002, keilmann2004, schliesser2005} can be combined with the advantages of photo-acoustic detection. This novel approach of \textit{photo-acoustic dual-frequency comb spectroscopy (PA-DCS)} enables the rapid and scan-free acquisition of absorption features with high resolution and precision via photo-acoustic dual-comb multi-heterodyne detection.

Figure~\ref{fig1}b illustrates the concept of PA-DCS. Similar to conventional DCS, two frequency combs are used in our demonstration whose optical frequency components $\nu_n^{(i)}$ are described by 
\begin{equation}
\nu_n^{(i)}=n \cdot f_\mathrm{rep}^{(i)} + \nu_\mathrm{0}^{(i)} \, , 
\end{equation}
$f_\mathrm{rep}^{(i)}$ and $\nu_\mathrm{0}^{(i)}$ denote the repetition rate (i.e. the comb line spacing) and the combs' optical offset frequencies, respectively. The index $i=1,2$ distinguishes the two combs, and $n=0,\pm 1, \pm 2, ..$ are the comb line indices. The combs' repetition rates and offsets differ only by small amounts $\Delta f_\mathrm{rep}=\left| f_\mathrm{rep}^{(1)} - f_\mathrm{rep}^{(2)}\right| \ll f_\mathrm{rep}^{(1,2)}$, and $\Delta\nu_0 = \left| \nu_\mathrm{0}^{(1)}-\nu_\mathrm{0}^{(2)} \right|\ll f_\mathrm{rep}^{(1,2)}$, so that pairs of optical comb lines $\nu_n^{(1)}$ and $\nu_n^{(2)}$ are only separated by acoustic frequencies. When both combs are optically combined, this can be interpreted as a \textit{single} frequency comb
\begin{equation}
\tilde{\nu}_n=n \cdot \tfrac{1}{2} \left( f_\mathrm{rep}^{(1)} + f_\mathrm{rep}^{(2)}\right)+ \tfrac{1}{2}\left( \nu_\mathrm{0}^{(1)}+\nu_\mathrm{0}^{(2)} \right)\, ,
\label{eq:singlecomb}
\end{equation}
whose optical lines are modulated in optical power according to $1+\cos{(2\pi f_n t)}$ with frequencies 
\begin{equation}
f_n=\left| \nu_n^{(1)}-\nu_n^{(2)} \right| = n\cdot \Delta f_\mathrm{rep} + \Delta \nu_0\, .
\label{eq:acousticcomb}
\end{equation}
Exposing the sample to the dual combs, it experiences periodic heating with frequency $f_n$ if light at the optical frequency $\tilde{\nu}_n$ is absorbed. The periodic heating will lead to the generation of heterodyne acoustic waves in function of the absorbed power. The superposition of all acoustic waves constitutes a $\Delta f_\mathrm{rep}^{-1}$-periodic interferogram that is detectable by a microphone or an equivalent transducer. In order to be detectable, all acoustic frequencies $f_n$ need to respect the bandwidth limitation of the transducer.

\begin{figure*}
    \includegraphics[width=\textwidth]{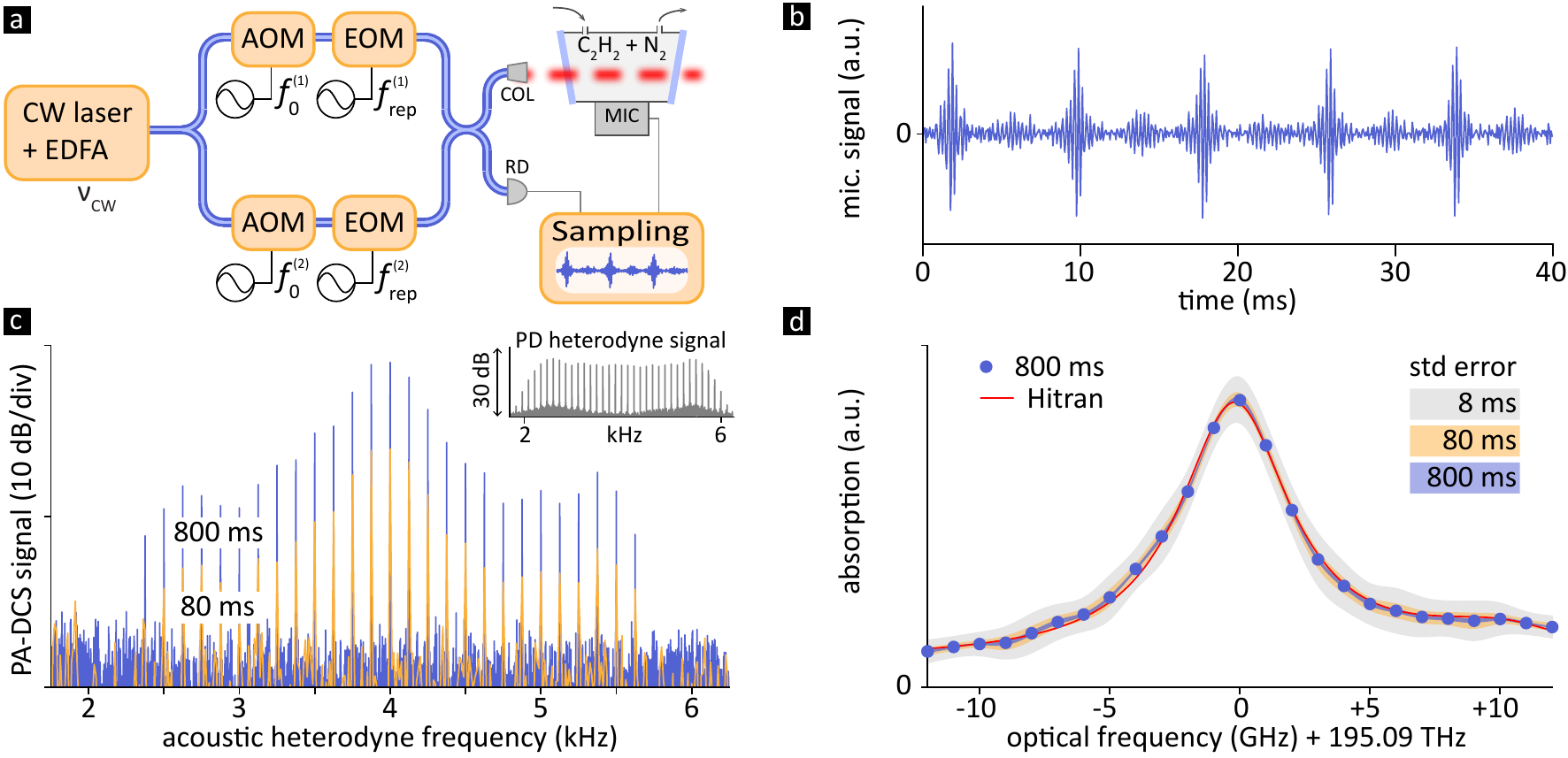}
    \caption{\textbf{Experimental setup and results.} 
        \textsf{\textbf{a}},~Experimental setup for the photo-acoustic detection of gaseous acetylene (C$_2$H$_2$). A tunable CW laser is amplified by an erbium-doped fiber amplifier (EDFA) and used as a common seed for the generation of two optical frequency combs via acousto-optic and electro-optic modulation (AOM, EOM). (COL:~free space collimator; RD:~reference (photo-)detector MIC:~low-noise MEMS microphone; see main text for more details). \textsf{\textbf{b}},~Acoustic multi-heterodyne signal recorded by the microphone (5~interferograms; after high-pass filtering) for an acetylene filled cell. \textsf{\textbf{c}},~Spectrum of the acoustic multi-heterodyne signal (acquisition time 80~ms and 800~ms). Inset: Multi-heterodyne reference spectrum as recorded by the reference (photo-)detector (over the same span). \textsf{\textbf{d}},~Acetylene absorption signature obtained after normalizing the acoustic multi-heterodyne spectrum by the reference spectrum and comparison to the HITRAN model.
        }        
    \label{fig2}
\end{figure*}
Key to our demonstration are dual-frequency combs with high mutual coherence that enable dense packing of the acoustic multi-heterodyne beatnotes $f_n$ within the microphone's bandwidth. Dual combs with high mutual coherence have been implemented in various ways based on mode-locked lasers or electro-optic modulation \cite{coddington2008, zolot2012,  millot2016, ideguchi2016, mehravar2016, zhao2016, link2017, hebert2018, chen2018, martin-mateos2018, gu2020}, and have also been extended to the infrared molecular fingerprint regime \cite{ycas2018,  muraviev2018, kayes2018, liao2018}. In this proof-of-concept demonstration, we use two near-infrared electro-optic combs (Figure~\ref{fig2}a). However, the concept can be readily applied to the mid-infrared domain.
In order to ensure high mutual coherence between both electro-optic combs, they are derived from a single continuous-wave (CW) laser with an optical frequency of $\nu_\mathrm{CW}=$195.09~THz (1536.71~nm). The amplified CW laser is split into two beams, each traversing first an acousto-optic modulator (AOM) where the laser frequencies are shifted by $f_0^{(1)}=80$~MHz and $f_0^{(1)}=80$~MHz + 4~kHz, respectively (i.e. $\nu_0^{(1,2)}=\nu_{CW}+f_0^{(1,2)}$), to create a relative comb offset of $\Delta \nu_0=4$~kHz. Next, each beam passes through an electro-optic
modulation (EOM) stage that includes one intensity and two phase modulators to generate a series of approximately 40 comb lines, spaced by  $f_\mathrm{rep}^{(1)}=1$~GHz and $f_\mathrm{rep}^{(2)}=f_\mathrm{rep}^{(1)} + 125$~Hz, respectively. In total, both combs deliver 20~mW of average power for photo-acoustic detection.
A simple aluminum tube (diameter 4~mm, length 10~mm) whose ends are sealed by two angled glass windows serves as the photo-acoustic sample chamber. Attached to the sidewall of the tube and connected through a small hole is an off-the-shelf micro-electro-mechanical system (MEMS) microphone with 20~kHz bandwidth. A battery-powered amplifier and digitizer is used to record the acoustic signals for a memory limited duration of up to 1~hour. The repetition rate difference of $\Delta f_\mathrm{rep}=125$~Hz was chosen so that all acoustic multi-heterodyne beatnotes would be within 4 to 6~kHz and well within the microphone's bandwidth. Given the combs' high mutual coherence, we note that more beatnotes (i.e. more optical sampling points) could readily be accommodated, particularly, when lowering $\Delta f_\mathrm{rep}$ or selecting a transducer with a larger bandwidth. The combs are sent through the sample tube in a multi-pass configuration. A small fraction of the comb light is sent to a photo-detector that provides a reference for normalization of the photo-acoustic signal and also enables an increase of the combs' mutual coherence in post-processing. In a wavelength regime where suitable photo-detectors may not be available, it could be replaced by photo-acoustic detection of black-body absorption. Both the microphone as well as the photo-detector signals are sampled. All modulation sources are synchronized to a 10~MHz frequency standard to ensure precise sampling and coherence in the acquisition process.

As a spectroscopic target we choose acetylene gas (C$_2$H$_2$) at atmospheric pressure and lab temperature as it provides precisely defined and interference-free absorption features. In a first experiment, the absorption cell is filled with acetylene gas, giving rise to the heterodyne acoustic interferogram signal shown in Figure~\ref{fig2}b. Figure~\ref{fig2}c shows two examples of the heterodyne acoustic spectra after Fourier-transformation \cite{virtanen2020} (PA-DCS signal) for acquisitions with a duration of 80~ms (10 interferograms) and 800~ms (100 interferograms), respectively. As expected, a longer acquisition time yields a higher SNR in the PA-DCS signal. In contrast to conventional DCS, photo-acoustic multi-heterodyne beatnotes are only generated in spectral regions that correspond to the presence of an absorber. Therefore, the number of photo-acoustic multi-heterodyne beatnotes is generally smaller than the number of comb lines. Although this does not allow for measurement of absolute absorption values without prior calibration, it avoids large (shot noise) background signals that can mask spectrally sparse or weak absorption features in conventional DCS \cite{newbury2010}.

In order to retrieve the true absorption profile, the acoustic multi-heterodyne beatnotes are normalized to account for the uneven spectral power envelope of the combs. Here, we accomplish this by dividing the PA-DCS signal (Figure~\ref{fig2}c) by the photo-detected heterodyne reference beatnotes (inset in Figure~\ref{fig2}c). The mapping of the acoustic to the optical frequency axis is described by Eqs.~\ref{eq:singlecomb} and \ref{eq:acousticcomb} (Note: this implies a scaling factor of $(f_\mathrm{rep}^{(1)}+f_\mathrm{rep}^{(2)})/(2\Delta f_\mathrm{rep})\approx 8 \times 10^6$ between acoustic and optical frequency axes). The resulting C$_2$H$_2$-absorption signature is shown and compared to the HITRAN-based prediction \cite{kochanov2016, gordon2017} in Figure~\ref{fig2}d:
Blue dots show the absorption retrieved from an 800~ms acquisition and shaded areas (grey, yellow, blue) represent the standard-error intervals for different durations of acquisition (8~ms, 80~ms and 800~ms). Excellent agreement between the HITRAN-prediction (red line) and the measured absorption profile is achieved, with the smallest residuals (below 3\% relative to peak absorption) observed with an 800~ms acquisition. The grey shaded area demonstrates that a fast, 8~ms acquisition (i.e. a single interferogram) is sufficient to retrieve the coarse features of the absorption profile.
The spectral resolution for each sampling point is given by the combs' absolute optical linewidth (here: $\sim$100~kHz), so that instrumental lineshape effects are negligible (resolution 5~orders of magnitude below the width of the absorption feature). Moreover, the frequency spacing of the sampling points (1.0000000675~GHz) is precisely defined by the mean repetition rate of the two combs (Eq.~\ref{eq:singlecomb}).

\begin{figure}
    \includegraphics[width=\columnwidth]{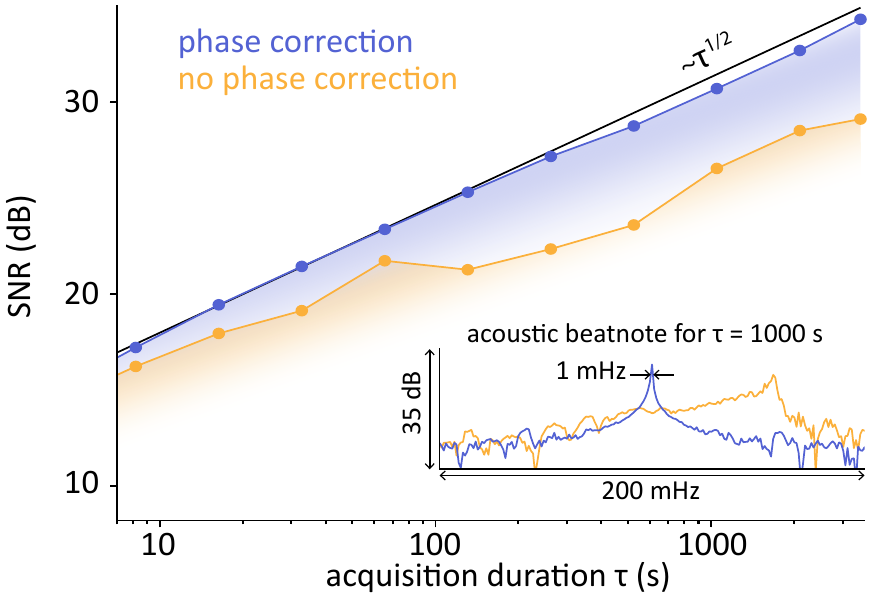}
    \caption{\textbf{Long-term acquisition.} 
        Signal-to-noise ratio (SNR) with (blue) and without (yellow) phase correction as a function of acquisition time $\tau$. The black line indicates the ideal case where the SNR increases proportionally to $\tau^{1/2}$. Inset: Spectrum of a heterodyne acoustic beatnote for $\tau=$1000~s with (blue) and without (yellow) phase correction.  
        }
        \label{fig3}
\end{figure}

Next, we investigate the extent to which even longer recordings of time $\tau$ can increase the SNR. To explore this, the cell is filled with N$_2$-diluted C$_2$H$_2$ with a concentration of 1\% and probed by combs centered at 1532.83~nm. Acquisitions of different duration are processed (similar to what is shown in Figure~\ref{fig2}) and the SNR of the highest acoustic beatnote (at 4~kHz) is determined as a function of $\tau$. Indeed, as Figure~\ref{fig3} shows, the SNR increases with $\tau$ (yellow trace), however, it markedly deviates from the $\tau^{1/2}$-scaling one would expect in the scenario with perfect noise-averaging. 
This deviation is due to small and slow length fluctuations in the non-common optical path of the combs that limits their mutual coherence on the time scale of few seconds or longer. These slow fluctuations manifest themselves as phase drifts in the multi-heterodyne beatnotes, which fortunately, can easily be tracked and corrected for numerically \cite{zolot2012, roy2012, ideguchi2014, burghoff2016, hebert2017,  zhu2018a, sterczewski2019}. Here, we extract the phase drift (one phase value for all heterodyne beats) from the reference heterodyne signal and, after low-pass filtering ($<0.1$~Hz), subtract it from the phase of the heterodyne acoustic beatnotes. This \textit{a-posteriori} phase-correction extends the effective mutual-coherence time of the combs by compensating for the slow path length fluctuations. As shown by the blue trace in Figure~\ref{fig3}, phase correction results in an increase of the SNR close to the ideal scaling (black line) up to the maximal recording duration of 1~hour. This result suggests that even longer recordings could be leveraged to further increase the signal to noise ratio. A small deviation from the ideal scaling is observed for acquisitions longer than 300~s and attributed to residual differential phase drifts between the heterodyne beatnotes, which could be addressed by tracking the phase of each beatnote separately.
To further illustrate the effect of phase correction, the inset in Figure~\ref{fig3} shows the spectrum of an individual heterodyne acoustic beatnote for a recording time of 1000~s. With phase correction applied (blue trace), a narrow 1~mHz linewidth heterodyne beatnote is detected.  Without phase correction (yellow trace) the drifting beatnote has a reduced SNR.
Generally, in photo-acoustic gas spectroscopy, the SNR depends on the used optical power, the absorption coefficient, the probing wavelength, the photo-acoustic cell design \cite{bijnen1996}, the microphone, the surrounding matter, environmental conditions (pressure, temperature) as well as the recording duration. In the current proof-of-concept configuration, based on the SNR in Figure~\ref{fig3}, we estimate a minimal detectable noise equivalent C$_2$H$_2$ concentration of 10~ppm for a recording time of 1000~s.\\

In conclusion, we have demonstrated photo-acoustic dual-frequency comb spectroscopy (PA-DCS) as a novel spectroscopic technique that can achieve high resolution, rapid acquisition and sensitive detection. Owing to its principle, PA-DCS can operate at any wavelength and across the large spectral bandwidth accessible to frequency combs \cite{schliesser2012, picque2019}. In the mid-infrared and other wavelength regimes where photo-detection is challenging, it can complement alternative DCS approaches, such as those based on optical field sampling or up-conversion \cite{kowligy2019, chen2020}. In future work, PA-DCS could be combined with high power quantum-cascade dual-laser combs \cite{villares2014, gianella2020}, cavity enhancement \cite{bernhardt2010}, multi-MHz bandwidth optical microphones \cite{rosenthal2011}, as well as opto-mechanical transducers \cite{gavartin2012, stiller2019, mason2019} to further increase sensitivity, spectral coverage and acquisition speed. As such, our demonstration generates new opportunities for rapid, sensitive broadband multi-species molecular analysis across all wavelengths of light.

\section*{}
\footnotesize{This work was supported by the Swiss National Science Foundation (00020\_182598), the Helmholtz Young Investigators Group VH-NG-1404 and the Canton of Neuchâtel.}

\bibliographystyle{unsrt}
\bibliography{arxiv}

\begin{thebibliography}{10}

\bibitem{kreuzer1971}
L.~B. Kreuzer and C.~K.~N. Patel.
\newblock Nitric {Oxide} {Air} {Pollution}: {Detection} by {Optoacoustic}
  {Spectroscopy}.
\newblock {\em Science}, 173(3991):45--47, July 1971.

\bibitem{rosencwaig1973}
A.~Rosencwaig.
\newblock Photoacoustic {Spectroscopy} of {Biological} {Materials}.
\newblock {\em Science}, 181(4100):657--658, August 1973.

\bibitem{patel1979}
C.~K.~N. Patel and A.~C. Tam.
\newblock Optoacoustic spectroscopy of liquids.
\newblock {\em Applied Physics Letters}, 34(7):467--470, April 1979.
\newblock Publisher: American Institute of Physics.

\bibitem{li2011}
Jingsong Li, Weidong Chen, and Benli Yu.
\newblock Recent {Progress} on {Infrared} {Photoacoustic} {Spectroscopy}
  {Techniques}.
\newblock {\em Applied Spectroscopy Reviews}, 46(6):440--471, August 2011.

\bibitem{cremer2016}
Johannes~W. Cremer, Klemens~M. Thaler, Christoph Haisch, and Ruth Signorell.
\newblock Photoacoustics of single laser-trapped nanodroplets for the direct
  observation of nanofocusing in aerosol photokinetics.
\newblock {\em Nature Communications}, 7(1):1--7, March 2016.
\newblock Number: 1 Publisher: Nature Publishing Group.

\bibitem{herpen2006}
M.~M. J. W.~van Herpen, A.~K.~Y. Ngai, S.~E. Bisson, J.~H.~P. Hackstein, E.~J.
  Woltering, and F.~J.~M. Harren.
\newblock Optical parametric oscillator-based photoacoustic detection of {CO} 2
  at 4.23 um allows real-time monitoring of the respiration of small insects.
\newblock {\em Applied Physics B}, 82(4):665--669, March 2006.
\newblock Company: Springer Distributor: Springer Institution: Springer Label:
  Springer Number: 4 Publisher: Springer-Verlag.

\bibitem{wang2003}
Xueding Wang, Yongjiang Pang, Geng Ku, Xueyi Xie, George Stoica, and Lihong~V.
  Wang.
\newblock Noninvasive laser-induced photoacoustic tomography for structural and
  functional in vivo imaging of the brain.
\newblock {\em Nature Biotechnology}, 21(7):803--806, July 2003.
\newblock Number: 7 Publisher: Nature Publishing Group.

\bibitem{kosterev2002}
A.~A. Kosterev, Yu~A. Bakhirkin, R.~F. Curl, and F.~K. Tittel.
\newblock Quartz-enhanced photoacoustic spectroscopy.
\newblock {\em Optics Letters}, 27(21):1902--1904, November 2002.
\newblock Publisher: Optical Society of America.

\bibitem{wu2017}
Hongpeng Wu, Lei Dong, Huadan Zheng, Yajun Yu, Weiguang Ma, Lei Zhang, Wangbao
  Yin, Liantuan Xiao, Suotang Jia, and Frank~K. Tittel.
\newblock Beat frequency quartz-enhanced photoacoustic spectroscopy for fast
  and calibration-free continuous trace-gas monitoring.
\newblock {\em Nature Communications}, 8(1):1--8, May 2017.
\newblock Number: 1 Publisher: Nature Publishing Group.

\bibitem{koskinen2007}
V.~Koskinen, J.~Fonsen, K.~Roth, and J.~Kauppinen.
\newblock Cantilever enhanced photoacoustic detection of carbon dioxide using a
  tunable diode laser source.
\newblock {\em Applied Physics B}, 86(3):451--454, February 2007.
\newblock Company: Springer Distributor: Springer Institution: Springer Label:
  Springer Number: 3 Publisher: Springer-Verlag.

\bibitem{peltola2013}
Jari Peltola, Markku Vainio, Tuomas Hieta, Juho Uotila, Sauli Sinisalo, Markus
  Metsälä, Mikael Siltanen, and Lauri Halonen.
\newblock High sensitivity trace gas detection by cantilever-enhanced
  photoacoustic spectroscopy using a mid-infrared continuous-wave optical
  parametric oscillator.
\newblock {\em Optics Express}, 21(8):10240--10250, April 2013.
\newblock Publisher: Optical Society of America.

\bibitem{spagnolo2012}
Vincenzo Spagnolo, Pietro Patimisco, Simone Borri, Gaetano Scamarcio, Bruce~E.
  Bernacki, and Jason Kriesel.
\newblock Part-per-trillion level {SF}$_{\textrm{6}}$ detection using a quartz
  enhanced photoacoustic spectroscopy-based sensor with single-mode
  fiber-coupled quantum cascade laser excitation.
\newblock {\em Optics Letters}, 37(21):4461--4463, November 2012.
\newblock Publisher: Optical Society of America.

\bibitem{tomberg2018}
Teemu Tomberg, Markku Vainio, Tuomas Hieta, and Lauri Halonen.
\newblock Sub-parts-per-trillion level sensitivity in trace gas detection by
  cantilever-enhanced photo-acoustic spectroscopy.
\newblock {\em Scientific Reports}, 8(1):1848, December 2018.

\bibitem{busse1978}
G~Busse and B~Bullemer.
\newblock Use of the opto-acoustic effect for rapid scan {Fourier}
  spectroscopy.
\newblock {\em Infrared Physics}, 18:631--634, 1978.

\bibitem{mikkonen2018}
Tommi Mikkonen, Caroline Amiot, Antti Aalto, Kim Patokoski, Goëry Genty, and
  Juha Toivonen.
\newblock Broadband cantilever-enhanced photoacoustic spectroscopy in the
  mid-{IR} using a supercontinuum.
\newblock {\em Optics Letters}, 43(20):5094--5097, October 2018.
\newblock Publisher: Optical Society of America.

\bibitem{sadiek2018}
Ibrahim Sadiek, Tommi Mikkonen, Markku Vainio, Juha Toivonen, and Aleksandra
  Foltynowicz.
\newblock Optical frequency comb photoacoustic spectroscopy.
\newblock {\em Physical Chemistry Chemical Physics}, 20(44):27849--27855, 2018.

\bibitem{karhu2019}
Juho Karhu, Teemu Tomberg, Francisco~Senna Vieira, Guillaume Genoud, Vesa
  Hänninen, Markku Vainio, Markus Metsälä, Tuomas Hieta, Steven Bell, and
  Lauri Halonen.
\newblock Broadband photoacoustic spectroscopy of {14CH4} with a high-power
  mid-infrared optical frequency comb.
\newblock {\em Optics Letters}, 44(5):1142--1145, March 2019.
\newblock Publisher: Optical Society of America.

\bibitem{maslowski2016}
Piotr Maslowski.
\newblock Surpassing the path-limited resolution of {Fourier}-transform
  spectrometry with frequency combs.
\newblock {\em Physical Review A}, 93(2), 2016.

\bibitem{coddington2016}
Ian Coddington, Nathan Newbury, and William Swann.
\newblock Dual-comb spectroscopy.
\newblock {\em Optica}, 3(4):414--426, April 2016.
\newblock Publisher: Optical Society of America.

\bibitem{picque2019}
Nathalie Picqué and Theodor~W. Hänsch.
\newblock Frequency comb spectroscopy.
\newblock {\em Nature Photonics}, 13(3):146--157, March 2019.
\newblock Number: 3 Publisher: Nature Publishing Group.

\bibitem{schiller2002}
S.~Schiller.
\newblock Spectrometry with frequency combs.
\newblock {\em Optics Letters}, 27(9):766--768, May 2002.
\newblock Publisher: Optical Society of America.

\bibitem{keilmann2004}
Fritz Keilmann, Christoph Gohle, and Ronald Holzwarth.
\newblock Time-domain mid-infrared frequency-comb spectrometer.
\newblock {\em Optics Letters}, 29(13):1542--1544, July 2004.
\newblock Publisher: Optical Society of America.

\bibitem{schliesser2005}
Albert Schliesser, Markus Brehm, Fritz Keilmann, and Daniel W. van~der Weide.
\newblock Frequency-comb infrared spectrometer for rapid, remote chemical
  sensing.
\newblock {\em Optics Express}, 13(22):9029--9038, October 2005.
\newblock Publisher: Optical Society of America.

\bibitem{coddington2008}
Ian Coddington.
\newblock Coherent {Multiheterodyne} {Spectroscopy} {Using} {Stabilized}
  {Optical} {Frequency} {Combs}.
\newblock {\em Physical Review Letters}, 100(1), 2008.

\bibitem{zolot2012}
A.~M. Zolot, F.~R. Giorgetta, E.~Baumann, J.~W. Nicholson, W.~C. Swann,
  I.~Coddington, and N.~R. Newbury.
\newblock Direct-comb molecular spectroscopy with accurate, resolved comb teeth
  over 43 {THz}.
\newblock {\em Optics Letters}, 37(4):638, February 2012.
\newblock publisher: Optical Society of America.

\bibitem{millot2016}
Guy Millot, Stéphane Pitois, Ming Yan, Tatevik Hovhannisyan, Abdelkrim
  Bendahmane, Theodor~W. Hänsch, and Nathalie Picqué.
\newblock Frequency-agile dual-comb spectroscopy.
\newblock {\em Nature Photonics}, 10(1):27--30, January 2016.
\newblock Number: 1 Publisher: Nature Publishing Group.

\bibitem{ideguchi2016}
Takuro Ideguchi, Tasuku Nakamura, Yohei Kobayashi, and Keisuke Goda.
\newblock Kerr-lens mode-locked bidirectional dual-comb ring laser for
  broadband dual-comb spectroscopy.
\newblock {\em Optica}, 3(7):748--753, July 2016.
\newblock Publisher: Optical Society of America.

\bibitem{mehravar2016}
S.~Mehravar, R.~A. Norwood, N.~Peyghambarian, and K.~Kieu.
\newblock Real-time dual-comb spectroscopy with a free-running bidirectionally
  mode-locked fiber laser.
\newblock {\em Applied Physics Letters}, 108(23):231104, June 2016.
\newblock Publisher: American Institute of Physics.

\bibitem{zhao2016}
Xin Zhao, Guoqing Hu, Bofeng Zhao, Cui Li, Yingling Pan, Ya~Liu, Takeshi Yasui,
  and Zheng Zheng.
\newblock Picometer-resolution dual-comb spectroscopy with a free-running fiber
  laser.
\newblock {\em Optics Express}, 24(19):21833--21845, September 2016.
\newblock Publisher: Optical Society of America.

\bibitem{link2017}
S.~M. Link, D.~J. H.~C. Maas, D.~Waldburger, and U.~Keller.
\newblock Dual-comb spectroscopy of water vapor with a free-running
  semiconductor disk laser.
\newblock {\em Science}, 356(6343):1164--1168, June 2017.
\newblock Publisher: American Association for the Advancement of Science
  Section: Reports.

\bibitem{hebert2018}
Nicolas~Bourbeau Hébert, David~G. Lancaster, Vincent Michaud-Belleau,
  George~Y. Chen, and Jérôme Genest.
\newblock Highly coherent free-running dual-comb chip platform.
\newblock {\em Optics Letters}, 43(8):1814--1817, April 2018.
\newblock Publisher: Optical Society of America.

\bibitem{chen2018}
Zaijun Chen, Ming Yan, Theodor~W. Hänsch, and Nathalie Picqué.
\newblock A phase-stable dual-comb interferometer.
\newblock {\em Nature Communications}, 9(1):1--7, August 2018.
\newblock Number: 1 Publisher: Nature Publishing Group.

\bibitem{martin-mateos2018}
Pedro Martín-Mateos, Borja Jerez, Pedro Largo-Izquierdo, and Pablo Acedo.
\newblock Frequency accurate coherent electro-optic dual-comb spectroscopy in
  real-time.
\newblock {\em Optics Express}, 26(8):9700--9713, April 2018.
\newblock Publisher: Optical Society of America.

\bibitem{gu2020}
Chenglin Gu, Zhong Zuo, Daping Luo, Zejiang Deng, Yang Liu, Minglie Hu, and
  Wenxue Li.
\newblock Passive coherent dual-comb spectroscopy based on optical-optical
  modulation with free running lasers.
\newblock {\em PhotoniX}, 1(1):7, March 2020.

\bibitem{ycas2018}
Gabriel Ycas, Fabrizio~R. Giorgetta, Esther Baumann, Ian Coddington, Daniel
  Herman, Scott~A. Diddams, and Nathan~R. Newbury.
\newblock High-coherence mid-infrared dual-comb spectroscopy spanning 2.6 to
  5.2 um.
\newblock {\em Nature Photonics}, 12(4):202--208, April 2018.
\newblock Number: 4 Publisher: Nature Publishing Group.

\bibitem{muraviev2018}
A.~V. Muraviev, V.~O. Smolski, Z.~E. Loparo, and K.~L. Vodopyanov.
\newblock Massively parallel sensing of trace molecules and their isotopologues
  with broadband subharmonic mid-infrared frequency combs.
\newblock {\em Nature Photonics}, 12(4):209--214, April 2018.
\newblock Number: 4 Publisher: Nature Publishing Group.

\bibitem{kayes2018}
M.~Imrul Kayes, Nurmemet Abdukerim, Alexandre Rekik, and Martin Rochette.
\newblock Free-running mode-locked laser based dual-comb spectroscopy.
\newblock {\em Optics Letters}, 43(23):5809--5812, December 2018.
\newblock Publisher: Optical Society of America.

\bibitem{liao2018}
Ruoyu Liao, Youjian Song, Wu~Liu, Haosen Shi, Lu~Chai, and Minglie Hu.
\newblock Dual-comb spectroscopy with a single free-running thulium-doped fiber
  laser.
\newblock {\em Optics Express}, 26(8):11046--11054, April 2018.
\newblock Publisher: Optical Society of America.

\bibitem{virtanen2020}
Pauli Virtanen, Ralf Gommers, Travis~E. Oliphant, Matt Haberland, Tyler Reddy,
  David Cournapeau, Evgeni Burovski, Pearu Peterson, Warren Weckesser, Jonathan
  Bright, Stéfan J. van~der Walt, Matthew Brett, Joshua Wilson, K.~Jarrod
  Millman, Nikolay Mayorov, Andrew R.~J. Nelson, Eric Jones, Robert Kern, Eric
  Larson, C.~J. Carey, İlhan Polat, Yu~Feng, Eric~W. Moore, Jake VanderPlas,
  Denis Laxalde, Josef Perktold, Robert Cimrman, Ian Henriksen, E.~A. Quintero,
  Charles~R. Harris, Anne~M. Archibald, Antônio~H. Ribeiro, Fabian Pedregosa,
  and Paul~van Mulbregt.
\newblock {SciPy} 1.0: fundamental algorithms for scientific computing in
  {Python}.
\newblock {\em Nature Methods}, 17(3):261--272, March 2020.
\newblock Number: 3 Publisher: Nature Publishing Group.

\bibitem{newbury2010}
Nathan~R. Newbury, Ian Coddington, and William Swann.
\newblock Sensitivity of coherent dual-comb spectroscopy.
\newblock {\em Optics Express}, 18(8):7929--7945, April 2010.
\newblock Publisher: Optical Society of America.

\bibitem{kochanov2016}
R.V. Kochanov, I.E. Gordon, L.S. Rothman, P.~Wcisło, C.~Hill, and J.S.
  Wilzewski.
\newblock {HITRAN} {Application} {Programming} {Interface} ({HAPI}): {A}
  comprehensive approach to working with spectroscopic data.
\newblock {\em Journal of Quantitative Spectroscopy and Radiative Transfer},
  177:15--30, July 2016.

\bibitem{gordon2017}
I.E. Gordon, L.S. Rothman, C.~Hill, R.V. Kochanov, Y.~Tan, P.F. Bernath,
  M.~Birk, V.~Boudon, A.~Campargue, K.V. Chance, B.J. Drouin, J.-M. Flaud, R.R.
  Gamache, J.T. Hodges, D.~Jacquemart, V.I. Perevalov, A.~Perrin, K.P. Shine,
  M.-A.H. Smith, J.~Tennyson, G.C. Toon, H.~Tran, V.G. Tyuterev, A.~Barbe, A.G.
  Császár, V.M. Devi, T.~Furtenbacher, J.J. Harrison, J.-M. Hartmann,
  A.~Jolly, T.J. Johnson, T.~Karman, I.~Kleiner, A.A. Kyuberis, J.~Loos, O.M.
  Lyulin, S.T. Massie, S.N. Mikhailenko, N.~Moazzen-Ahmadi, H.S.P. Müller,
  O.V. Naumenko, A.V. Nikitin, O.L. Polyansky, M.~Rey, M.~Rotger, S.W. Sharpe,
  K.~Sung, E.~Starikova, S.A. Tashkun, J.~Vander Auwera, G.~Wagner,
  J.~Wilzewski, P.~Wcisło, S.~Yu, and E.J. Zak.
\newblock The {HITRAN2016} molecular spectroscopic database.
\newblock {\em Journal of Quantitative Spectroscopy and Radiative Transfer},
  203:3--69, December 2017.

\bibitem{roy2012}
Julien Roy, Jean-Daniel Deschênes, Simon Potvin, and Jérôme Genest.
\newblock Continuous real-time correction and averaging for frequency comb
  interferometry.
\newblock {\em Optics Express}, 20(20):21932--21939, September 2012.
\newblock Publisher: Optical Society of America.

\bibitem{ideguchi2014}
Takuro Ideguchi, Antonin Poisson, Guy Guelachvili, Nathalie Picqué, and
  Theodor~W. Hänsch.
\newblock Adaptive real-time dual-comb spectroscopy.
\newblock {\em Nature Communications}, 5(1):1--8, February 2014.
\newblock Number: 1 Publisher: Nature Publishing Group.

\bibitem{burghoff2016}
David Burghoff, Yang Yang, and Qing Hu.
\newblock Computational multiheterodyne spectroscopy.
\newblock {\em Science Advances}, 2(11):e1601227, November 2016.
\newblock Publisher: American Association for the Advancement of Science
  Section: Research Article.

\bibitem{hebert2017}
Nicolas~Bourbeau Hébert, Jérôme Genest, Jean-Daniel Deschênes, Hugo
  Bergeron, George~Y. Chen, Champak Khurmi, and David~G. Lancaster.
\newblock Self-corrected chip-based dual-comb spectrometer.
\newblock {\em Optics Express}, 25(7):8168--8179, April 2017.
\newblock Publisher: Optical Society of America.

\bibitem{zhu2018a}
Zebin Zhu, Kai Ni, Qian Zhou, and Guanhao Wu.
\newblock Digital correction method for realizing a phase-stable dual-comb
  interferometer.
\newblock {\em Optics Express}, 26(13):16813--16823, June 2018.
\newblock Publisher: Optical Society of America.

\bibitem{sterczewski2019}
Lukasz~A. Sterczewski, Lukasz~A. Sterczewski, Lukasz~A. Sterczewski, Jonas
  Westberg, and Gerard Wysocki.
\newblock Computational coherent averaging for free-running dual-comb
  spectroscopy.
\newblock {\em Optics Express}, 27(17):23875--23893, August 2019.
\newblock Publisher: Optical Society of America.

\bibitem{bijnen1996}
F.~G.~C. Bijnen, J.~Reuss, and F.~J.~M. Harren.
\newblock Geometrical optimization of a longitudinal resonant photoacoustic
  cell for sensitive and fast trace gas detection.
\newblock {\em Review of Scientific Instruments}, 67(8):2914--2923, August
  1996.
\newblock Publisher: American Institute of Physics.

\bibitem{schliesser2012}
Albert Schliesser, Nathalie Picqué, and Theodor~W. Hänsch.
\newblock Mid-infrared frequency combs.
\newblock {\em Nature Photonics}, 6(7):440--449, July 2012.
\newblock Number: 7 Publisher: Nature Publishing Group.

\bibitem{kowligy2019}
Abijith~S. Kowligy, Henry Timmers, Alexander~J. Lind, Ugaitz Elu, Flavio~C.
  Cruz, Peter~G. Schunemann, Jens Biegert, and Scott~A. Diddams.
\newblock Infrared electric field sampled frequency comb spectroscopy.
\newblock {\em Science Advances}, 5(6):8794, June 2019.
\newblock Publisher: American Association for the Advancement of Science
  Section: Research Article.

\bibitem{chen2020}
Zaijun Chen, Theodor~W. Hänsch, and Nathalie Picqué.
\newblock Upconversion mid-infrared dual-comb spectroscopy.
\newblock {\em arXiv:2003.06930 [physics]}, March 2020.
\newblock arXiv: 2003.06930.

\bibitem{villares2014}
Gustavo Villares, Andreas Hugi, Stéphane Blaser, and Jérôme Faist.
\newblock Dual-comb spectroscopy based on quantum-cascade-laser frequency
  combs.
\newblock {\em Nature Communications}, 5(1):1--9, October 2014.
\newblock Number: 1 Publisher: Nature Publishing Group.

\bibitem{gianella2020}
Michele Gianella, Akshay Nataraj, Béla Tuzson, Pierre Jouy, Filippos
  Kapsalidis, Mattias Beck, Markus Mangold, Andreas Hugi, Jérôme Faist, and
  Lukas Emmenegger.
\newblock High-resolution and gapless dual comb spectroscopy with current-tuned
  quantum cascade lasers.
\newblock {\em Optics Express}, 28(5):6197--6208, March 2020.
\newblock Publisher: Optical Society of America.

\bibitem{bernhardt2010}
Birgitta Bernhardt, Akira Ozawa, Patrick Jacquet, Marion Jacquey, Yohei
  Kobayashi, Thomas Udem, Ronald Holzwarth, Guy Guelachvili, Theodor~W.
  Hänsch, and Nathalie Picqué.
\newblock Cavity-enhanced dual-comb spectroscopy.
\newblock {\em Nature Photonics}, 4(1):55--57, January 2010.
\newblock Number: 1 Publisher: Nature Publishing Group.

\bibitem{rosenthal2011}
Amir Rosenthal, Daniel Razansky, and Vasilis Ntziachristos.
\newblock High-sensitivity compact ultrasonic detector based on a
  pi-phase-shifted fiber {Bragg} grating.
\newblock {\em Optics Letters}, 36(10):1833--1835, May 2011.
\newblock Publisher: Optical Society of America.

\bibitem{gavartin2012}
E.~Gavartin, P.~Verlot, and T.~J. Kippenberg.
\newblock A hybrid on-chip optomechanical transducer for ultrasensitive force
  measurements.
\newblock {\em Nature Nanotechnology}, 7(8):509--514, August 2012.
\newblock Number: 8 Publisher: Nature Publishing Group.

\bibitem{stiller2019}
Birgit Stiller, Paulo Dainese, and Ewold Verhagen.
\newblock Optoacoustics—{Advances} in high-frequency optomechanics and
  {Brillouin} scattering.
\newblock {\em APL Photonics}, 4:110401, May 2019.
\newblock Publisher: American Institute of Physics.

\bibitem{mason2019}
David Mason, Junxin Chen, Massimiliano Rossi, Yeghishe Tsaturyan, and Albert
  Schliesser.
\newblock Continuous force and displacement measurement below the standard
  quantum limit.
\newblock {\em Nature Physics}, 15(8):745--749, August 2019.
\newblock Number: 8 Publisher: Nature Publishing Group.

\end{thebibliography}

\end{document}